\documentclass[12pt,a4paper]{article}

\usepackage{a4wide,amssymb}

\newcommand\eqref[1]{(\ref{#1})}
\newcommand\II{I\kern-0.1em I}
\newcommand\inv{^{\raise.15ex\hbox{${\scriptscriptstyle -}$}\kern-.05em 1}} 
\newcommand\grad[2][]{\mathop{\nabla_{\!#2}^{#1}}\nolimits}	
\newcommand\hatgrad[2][]{\mathop{\widehat\nabla_{\!#2}^{#1}}\nolimits}
\newcommand\one{\hbox{1\kern-.25em l}}			
\newcommand\con{\mathbin{\lrcorner}}			
\newcommand\st{{}^*\kern-.14em}				
\newcommand\CA{\mathcal{A}}
\newcommand\CB{\mathcal{B}}
\newcommand\CC{\mathcal{C}}
\newcommand\CE{\mathcal{E}}
\newcommand\CF{\mathcal{F}}
\newcommand\CG{\mathcal{G}}
\newcommand\CH{\mathcal{H}}

\newcommand\CL{\mathcal{L}}
\newcommand\CM{\mathcal{M}}
\newcommand\CN{\mathcal{N}}
\newcommand\bbC{\mathbb{C}}
\newcommand\bbG{\mathbb{G}}
\newcommand\eps{\epsilon}

\newcommand\set[1]{\mathbb{#1}}				
\newcommand\Z{\set{Z}}
\newcommand\T{\set{T}}
\newcommand\group[1]{\mathop{\kern0pt\mathrm{#1}}\nolimits}	
\newcommand\U{\group{U}}				
\newcommand\SU{\group{SU}}	
\renewcommand\O{\group{O}}	
\newcommand\SO{\group{SO}}	
\newcommand\SL{\group{SL}}
\newcommand\E{\group{E}}
\newcommand\lie[1]{\mathop{\kern0pt\mathit{#1}}\nolimits}	

\newcommand\opname[1]{\mathop{\kern0pt\mathrm{#1}}\nolimits}	
\newcommand\Tr{\opname{Tr}}				
\newcommand\tr{\opname{tr}}				
\renewcommand\det{\opname{det}}				
\newcommand\diag{\opname{diag}}				
\newcommand\e{\opname{e}}				
\newcommand\ch{\opname{ch}}				
\makeatletter
\renewcommand\section{\@startsection{section}{1}{\z@}%
                                    {-6ex \@plus -1ex \@minus -.2ex}%
                                    {2.5ex \@plus.2ex}%
                                    {\normalfont\large\scshape\centering}}                                   
\renewcommand\subsection{\@startsection{subsection}{2}{\z@}%
                                       {5ex \@plus -1ex \@minus -.2ex}%
                                       {1.5ex \@plus.2ex}%
                                       {\normalfont\normalsize\scshape}}
\renewcommand\abstract{\section*{\abstractname}}                                   
\newcommand\ack{\section*{\ackname}}                                   
\newcommand\ackname{Acknowledgements}
\renewcommand\appendix{\par
  \setcounter{section}{0}%
  \setcounter{subsection}{0}%
  \renewcommand\thesection{\appendixname~\@Alph\c@section}}
\renewcommand\@seccntformat[1]{\csname the#1\endcsname.\quad}
\long\def\@makecaption#1#2{%
  \vskip\abovecaptionskip
  \sbox\@tempboxa{\textsc{#1}: #2}%
  \ifdim \wd\@tempboxa >\hsize
    \textsc{#1}: #2\par
  \else
    \global \@minipagefalse
    \hb@xt@\hsize{\hfil\box\@tempboxa\hfil}%
  \fi
  \vskip\belowcaptionskip}
\makeatother					
\begin{document}

%
%

\begin{flushright}
\textsc{spin-98/3\\ hep-th/9810219\\ Oktober 1998}
\end{flushright}
\vskip1cm

\begin{center}

{\LARGE\scshape Gauge Bundles and Born-Infeld \\ on the Noncommutative Torus \par}
\vskip25mm

\textsc{Christiaan Hofman} and 
\textsc{Erik Verlinde}\\[3mm]

\textit{Spinoza Institute \\ \textnormal{and}\\ Institute for Theoretical Physics\\
        University of Utrecht, 3584 CE Utrecht}\\[2mm]

\texttt{hofman@phys.uu.nl, verlinde@phys.uu.nl}\\
 
\end{center}
\vskip3cm

\abstract

In this paper, we describe non-abelian gauge bundles with magnetic and electric fluxes 
on higher dimensional noncommutative tori. We give an explicit construction of a large 
class of bundles with nonzero magnetic 't Hooft fluxes. We discuss Morita equivalence 
between these bundles. The action of the duality is worked out in detail for the 
four-torus. As an application, we discuss Born-Infeld on this torus, as a description 
of compactified string theory. We show that the resulting theory, including the 
fluctuations, is manifestly invariant under the T-duality group $\SO(4,4;\Z)$. The 
U-duality invariant BPS mass-formula is discussed shortly. We comment on a discrepancy 
of this result with that of a recent calculation. 

\newpage
%
%

\section{Introduction}

In the last year, noncommutative geometry, and especially the noncommutative torus, 
got exciting new applications in compactifications of M-theory. This started with 
\cite{codo}, where compactification of M-theory on the noncommutative two-torus was 
studied. Many more discussions of M- and string theory compactifcations on these 
geometries followed, for example 
\cite{howu,howu2,dohu,li,cas,sch,kaok,chkr,arar,ho,mozu,bmz,brmo,hv}. The results of 
the last two papers are slightly different from ours. 

The noncommutative torus basically is a flat compact space, where, in contrast to the 
`classical' torus, the flat coordinates have nonzero commutation relations. These new 
compactifications were identified as string compactifications with nonzero $B$-flux
\cite{codo,dohu}. A nice property of the noncommutative torus, is the appearance of 
an Morita equivalence, which is a mathematical equivalence between gauge bundles on 
different tori. It was noted that in the connection to string theory Morita equivalence
expresses T-duality \cite{codo,sch}. 

In the papers cited above, mainly the two-torus was considered. In this paper we 
describe gauge theories on the noncommutative torus in higher dimensions. A 
different, more abstract approach to this was also taken in \cite{sch}. There it 
was also found that for the $d$-torus, the Morita equivalences form an $\SO(d,d;\Z)$ 
group, which is precisely the group needed for the identification as T-duality. 
We also give an explicit construction for a large class of gauge bundles with nonzero 
magnetic fluxes. Morita equivalence will come out automatically in this construction. 
We also include the electric variables in the discussion. 
For the case of the four-torus, we discuss these matters in some more detail.

String theory compactified on a four-torus, can contain several wrapped D-branes. 
The dynamics of D0-branes on the four-torus is known to be described by a gauge 
theory on the dual four-torus \cite{bfss,tay}. The resulting four-dimensional gauge 
theory can be viewed as the world-volume theory of the D4-brane. As is well-known,
the world-volume theory of a D-brane is given by a Born-Infeld type theory \cite{lei}.
Therefore at low energies, the system of D0-branes should be described by this 
Born-Infeld theory. The D4-brane system may contain bound states of wrapped D2-branes, 
D0-branes, and fundamental strings, as well as momentum. These different charges 
manifest themselves in the gauge theory as the different fluxes. So the different 
charge sectors of the Type \II{A} string theory are identified with different 
topological sectors in the gauge theory. The charges and the corresponding 
fluxes are presented in table \ref{tb:charges}. 

\begin{table}[h]
\begin{center}
\begin{tabular}{|c|c|c|c|}
\hline
charge   & D0-brane & D4-brane & Yang-Mills                      \\ \hline 
$N$      & D0       & D4       & rank                            \\
$M_{ij}$ & D2       & D2       & $\int\!\Tr F_{ij}$              \\
$k$      & D4       & D0       & $\int\!\Tr\frac{1}{2}F\wedge F$ \\
$n^i$    & momentum & winding  & $\int\!\Tr E^i$                 \\
$m_i$    & winding  & momentum & $\int\!\Tr P_i$                 \\ \hline
\end{tabular}
\caption{The different charges from the point of view of the D0-branes, the T-dual 
D4-branes, and the interpretation as fluxes in the gauge theory.}
\label{tb:charges}
\end{center}
\end{table}

In \cite{codo}, it was argued that in the presence of a constant background 
$B$-field, the dual torus on which the gauge theory lives should be replaced 
by a noncommutative torus. The noncommutative parameter $\theta$ should be 
identified with this constant $B$-field. On the other hand, the moduli on 
the T-dual D4-brane side, are modified as they should be given by the T-dual 
expression. Especially, the metric is different from the metric on dual 
momentum torus, which is just the inverse metric. So it seems that the 
theory of the D0-branes is different from the effective world-volume theory 
on the D4-brane. The two descriptions should however be closely related. 
So we still expect that the system should be described by a gauge theory of 
Born-Infeld type.

This paper is a generalization to higher dimensional noncommutative tori of 
an earlier paper \cite{hv}. But the main focus of this paper will be the study 
of general gauge bundles on the noncommutative tori in higher dimensions. 

In section \ref{nctorus}, we give a detailed description of gauge bundles on the 
noncommutative torus. We give explicit representations of the gauge field on the 
noncommutative $d$-torus for situations with magnetic and electric fluxes. We also 
discuss flux quantization on the noncommutative torus. We go into some depth 
concerning the electric flux quantization, because of the confusion in the recent 
literature. The action of duality transformations on the fluxes is discussed in 
more detail in four dimensions. 

In section \ref{ncbi}, the Born-Infeld theory living on the noncommutative 
four-torus will be introduced. The Hamiltonian of this theory will be fixed by 
requiring to have the same dependence on the global zero-modes, as this was already 
correctly described in the naive generalization. We discuss a \emph{manifest} 
invariance of the Hamiltonian under the full T-duality group $\SO(4,4;\Z)$.

Section \ref{concl} concludes with some discussion on what we found.

\section{The Noncommutative Torus}
\label{nctorus}

In this section we describe the noncommutative torus in some detail. 
We are mainly interested in an explicit description of gauge bundles. 
As we will have in mind the gauge theory on the noncommutative torus, 
the description will be in terms of the gauge connection, and not the 
sections of the bundles. We start with some generalities on 
noncommutative tori and bundles on them. Then we describe in detail 
the abelian and non-abelian gauge bundles.

\subsection{Noncommutative Geometry and Gauge Bundles}

It is quite often convenient to describe a geometry in terms of the (complex) 
functions on the manifold. As we can multiply and sum up these functions, they 
form an algebra. In the case of ordinary geometry, this algebra is of course 
always commutative. Usually one also wants to define an (Hermitian) inner 
product on the space of functions. The algebra then becomes what is called a 
$C^*$-algebra. It turns out that there is a one-to-one correspondence 
between commutative $C^*$-algebras and classical geometries. 

In noncommutative geometry, one starts from a $C^*$-algebra, but the demand 
that it should be a commutative algebra is dropped. This turns out to give  a 
quite general framework to study geometries which do not have a simple local 
description, such as fractal geometries.

Really the simplest example of a noncommutative geometry is the noncommutative 
torus. The classical torus is described in terms of the set of single valued 
functions. These are of course generated by the Fourier modes $U_i=\e^{2\pi i x_i}$. 
To define the noncommutative torus in $d$ dimensions, we start from the 
functions, which again are taken to be generated by Fourier modes. We will 
also denote these modes by $U_i$. Obviously, the algebra of functions is 
generated by these $d$ operators. Although in the classical case these modes 
obviously commute, for the noncommutative torus we assume that they do not 
commute, but have commutation relations of the form 
\begin{equation}\label{commU}
U_jU_k = \e^{2\pi i\theta^{jk}}U_kU_j,\qquad\mbox{or}\qquad
[x^j,x^k] = \frac{\theta^{jk}}{2\pi i},
\end{equation} 
where $\theta_{ij}$ is a constant anti-symmetric tensor. It is called the 
noncommutative parameter, as it characterizes the noncommutativeness of the 
torus. The noncommutative torus defined by this relation is denoted $\T^d_\theta$. 
Note that the above commutation relation can also be expressed in terms of a 
commutation relation between the local coordinates $x_i$ as 
$[x_j,x_k]=2\pi i\theta^{jk}$, which is essentially the commutation relation 
on a quantum phase space. Therefore the noncommutative torus is often referred 
to as the quantum torus. We will not use this name here, as we want to see the 
noncommutative geometry really as a classical geometry.

Except for the functions on the torus, we will also need a set of derivations. 
These will effectively replace the ordinary derivative operators on the torus, 
and can be considered as such. They are defined through the commutation relation 
with the algebra of functions. It is of obviously enough to give the commutation 
relations with the coordinates $x^k$ 
\begin{equation}
[\partial_j,x^k] = \delta_j^k.
\end{equation}  
Note that this is taken over without change from the classical situation. 
Also, remark that these derivations commute among themselves. This property 
will be very convenient, and allows one to consistently define differential 
forms on the noncommutative torus \cite{con,sch}. 

Another important ingredient for us is the definition of bundles on the torus. 
Also these bundles are in noncommutative geometry defined in terms of the sections. 
This generalizes the trivial ($\U(1)$) bundle, which has the functions as sections, 
and is therefore already defined above. An important property of these sections is 
that they can be multiplied (on the right) by a function. This makes them into 
(right-) modules of the $C^*$-algebra of functions. This characterization is the 
defining property of bundles on noncommutative geometry. More important for us is 
some extra structure on these bundles, namely the gauge connection. A connection on 
a bundle in noncommutative geometry is defined through the same requirement as in 
ordinary geometry, namely it has to be a derivation. This can be stated as 
\begin{equation}\label{gradper}
\grad{j}\bigl(\psi f\bigr) = \bigl(\grad{j}\psi\bigr)f + \psi\partial_j f,
\end{equation}
where $\psi$ is a section of the bundle and $f$ is any function on the 
noncommutative torus. 

Like for bundles on ordinary geometry, any connection can be written in terms 
of the derivations we defined above and a gauge field as 
$\grad{j}=\partial_j+i\CA_j$. Now the derivation condition above states that the 
gauge field $\CA$ should commute with all functions on the torus. Usually, this 
just says that it is a local function on the torus. In the noncommutative case, 
this is more subtle as we know that the functions do not commute among 
themselves. Therefore we first have to find a set of modified coordinates which 
commute with the coordinates $x^i$. It is easily seen that they are given by 
\begin{equation}\label{tildeU}
\tilde x^i = x^i+\frac{i\theta^{ij}}{2\pi}\partial_j.
\end{equation}
The derivative corrections are chosen such that the translations which the 
functions $x_i$ generate are cancelled. Analogously to the function modes $U_i$, we 
introduce modes $\widetilde U_i=\e^{2\pi i\tilde x^i}$. The gauge field should thus 
be functions of the $\tilde x^i$. In general, they will take of course values in the
Lie-algebra of the gauge group. 

Note that the modes $\tilde x^i$ satisfy commutation relations that are similar 
to that of the coordinates $x^i$, but with opposite noncommutative parameter 
\begin{equation}\label{commUtil}
[\tilde x^i,\tilde x^j]=-\frac{\theta^{ij}}{2\pi i}.
\end{equation}
The modes $\widetilde U_i$ therefore generate a $C^*$-algebra with parameter $-\theta^{ij}$. 

An essential notion for gauge bundles are local gauge transformations. 
These are given by gauge group-valued functions which act only fiber-wise. 
In noncommutative geometry, these should therefore be functions also of the 
$\tilde x^i$. This is also necessary for the gauge field to remain a function 
of the $\tilde x^i$  after a gauge transformation. 

To define a definite gauge bundle via its sections over the torus, we have 
to give periodicity conditions. This means that a translation, generated by 
$\e^{2\pi i\partial_i}$, acting on a section is equal to a gauge 
transformation $\Omega(\widetilde U)$. We can restate this by introducing a 
set of modified translation operators for the sections 
\begin{equation}
T_i=\e^{\partial_i}\Omega_i(\tilde x)\inv. 
\end{equation}
The allowed sections of the bundle are then simply the invariants of these 
operators, while the gauge connection should be invariant under conjugation by 
these operators. It turns out that all modules of the algebra of functions, 
and hence all bundles on the noncommutative torus, can be characterized in 
this way. The translation operators $T_i$ should commute among themselves
\footnote{In general this is only true if there is a fundamental bundle, 
otherwise they may commute up to elements of the center of the gauge group. 
In this paper, we assume that a fundamental bundle does exist. This is because 
we have in mind a relation to D-branes in string theory, and in these systems 
the endpoints of fundamental strings are fundamentals.}; this is equivalent 
with the cocycle condition on the $\Omega_i$. 

For example, for the trivial abelian gauge bundle, so all magnetic and higher 
fluxes are zero, the $\Omega_i$ are all trivial. It is easily seen that the 
sections are just the functions, generated by the $U_i$.
A simple connection on this bundle can also be given just by the derivations 
\begin{equation}\label{grad0ab}
\grad[0]{i}=\partial_i,
\end{equation}
which obviously commutes with the translation operators $T_j$. The gauge 
connection should be a function of the $\widetilde U_j=\e^{2\pi i\tilde x^i}$,
which commutes with the translation operators. It is then easily seen that it 
can be given as a power series in these modes 
\begin{equation}\label{conabel}
\CA_i = \sum_{p\in\Z^d}\CA_{i,p}\e^{2\pi ip_j\tilde x^j}.
\end{equation}
The gauge field should also satisfy a reality condition, which translates because 
of the choice above to $\CA_{j,p}^\dagger=\CA_{j,-p}$.

Trivial $\U(N)$ bundles are also easily constructed like this, by simply letting 
$\CA_j$ take values in the corresponding Lie-algebra.

\subsection{Non-Abelian Gauge Bundles}

We shall now turn to the description of non-abelian gauge bundles on the 
noncommutative torus. We also describe bundles with nontrivial fluxes. We 
assume here that the torus has even dimension $d=2g$. We study gauge bundle 
with gauge groups $\U(N)$ for arbitrary $N$, and magnetic fluxes $M$. There 
will be some restrictions on these fluxes. We do not know if these restrictions 
we will meet must be imposed in general, or if there are situations possible 
with more general fluxes. All the higher fluxes, such as the instanton number, 
will be, in the situations wer are able to describe, completely determined by 
the data $N$ and $M$. To be precise, the `integral Chern characters'
\footnote{We will see shortly that this is not the true Chern character in 
the noncommutative case, but it is closely related. In fact, what we mean here 
is the quantity we refer to as $\mu(F)$.} $\ch_k(F)$ for $k\geq1$ 
are given by 
\begin{equation}
\ch_k(F)=\frac{1}{k!}\Tr F^k = \frac{M^k}{k!N^{k-1}}.
\end{equation}

To define a particular $\U(N)$ bundle with the above magnetic fluxes $M$ can 
be defined through the translation operators $T_i$, defining the periodic 
boundary conditions in the different directions. Bundles with magnetic fluxes 
can be obtained by including $\SU(N)$ rotations in these boundary conditions. 
These $\SU(N)$ matrices, which we denote $V_i$, have to satisfy the commutation 
relations 
\begin{equation}
V_iV_j=\e^{2\pi iM_{ij}/N}V_jV_i,
\end{equation} 
where $M_{ij}$ is the integral matrix of fluxes. These matrices can only be 
constructed in certain cases, that is when $N$ is big enough to support them. 
An explicit construction of these matrices was worked out in \cite{geba}, and is 
discussed in \ref{twist}. There also these restrictions are discussed. Using these 
matrices, appropriate boundary conditions can be defined through the translation 
operators 
\begin{equation}
T_i=\e^{\partial_i}V_i\e^{2\pi i\alpha_{ij}\tilde x^j},
\end{equation}
where $\alpha$ is an appropriate matrix of abelian fluxes. It can be found 
from the requirement that the $T_i$ must commute. The solution to this 
constraint is not unique, but different solutions are gauge equivalent. 

We now turn to the explicit form the gauge field must take. The connection 
should satisfy the correct commutation relations with the Fourier modes 
$U_i$, as given in \eqref{gradper}, and it should commute with the translation 
operators $T_i$. Therefore we first construct the algebra of operators which 
commute with both the $T_i$ and the $U_i$. These then will be the modes from 
which the gauge field is constructed. This means that $\CA$ will be given 
by an expansion in these operators. They are the non-abelian generalization 
of the abelian operators $\widetilde U_i$. 

To write down the modes of the gauge field in the adjoint, we first introduce an 
integral matrix $\kappa$ which is as close the the inverse of $M$ as possible. 
To define it, we first introduce the matrix $n=\diag(n_1\one,\cdots,n_g\one)$, 
where $n_i=\gcd(M_i,N)$. Here we block-diagonalized $M$, 
$M=\diag(M_1\eps,\cdots,M_g\eps)$, and $\one$ is the two-by-two unit matrix. 
Now $\kappa$ and another integral matrix $\lambda$ are chosen such that 
\begin{equation}\label{mknlrel}
M\kappa + N\lambda = n.
\end{equation}
Note that in the basis where $M$ and $n$ are block-diagonal, also these two 
matrices are, $\kappa=\diag(\kappa_1\eps,\cdots,\kappa_g\eps)$ 
and $\lambda=\diag(\lambda_1\one,\cdots,\lambda_g\one)$. Note that then all 
four matrices we introduced commute. 
We define $\nu=Nn\inv$ and $\mu=Mn\inv$, which both are integral matrices.
In terms of these matrices, \eqref{mknlrel} can be written 
\begin{equation}
\mu\kappa+\nu\lambda=\one.
\end{equation}
With the matrix $\kappa$ at hand we can write the modes, which we shall denote 
$Z_i$, as 
\begin{equation}
Z_i=\e^{2\pi i\beta_{ij}\tilde x^j}\prod_jV_j^{\kappa_{ij}},
\end{equation}
where the matrix $\beta$ can be found from the requirement that $Z_i$ are inert 
under conjugation by the $T_i$ (this means that they are global sections of the 
gauge bundle). These modes satisfy commutation relations 
\begin{equation}\label{commZ}
Z_iZ_j = \e^{-2\pi i\hat\theta_{ij}}Z_jZ_i,
\qquad\mbox{where}\qquad
\hat\theta = (\kappa+\lambda\theta)\bigl(\nu-\mu\theta\bigr)\inv.
\end{equation}

The linear connection can easily be found from the representations of $T_i$ 
and $Z_i$, when we use the ansatz 
\begin{equation}
\exp(\grad[0]{i})=T_i\prod_jZ_j^{\mu_{ji}}.
\end{equation}
In this expression, the presence of the $T_i$ automatically guarantees the 
correct derivation condition \eqref{gradper}. The exponents of the $Z_i$ in 
this expression are precisely such that the $\SU(N)$ contribution (the part 
involving the $V_i$)). Therefore, we can solve for $\grad[0]{i}$ by taking the 
formal logarithm. 

The zero-mode of the field strength is then found from the linear connection 
\begin{equation}
F^0=\frac{1}{2\pi i}[\grad[0]{},\grad[0]{}]=(N-M\theta)\inv MI.
\end{equation}
Note that the precise choice for $\alpha$ drops out, 
therefore this field strength is gauge invariant, which it should be as it is 
an abelian connection. The complete connection can now be written as 
$\grad{i}=\grad[0]{i}+i\CA_i(Z)$, where $\CA_i$ can has an expansion in the 
non-abelian modes $Z_i$, analogous to \eqref{conabel}.

\subsection{Chern Character}

We now review the known expression of the Chern character on the noncommutative torus. 
For a more detailed discussion, see the book of Connes \cite{con}. 

The Chern character of a gauge bundle on the noncommutative torus is the element of 
the even cohomology 
\begin{equation}
\ch(F)=\Tr_\theta \e^{F},
\end{equation}
where $\Tr_\theta$ is the natural trace on the noncommutative torus. The element 
above should be interpreted as an element of the cohomology.

Any gauge bundle on the noncommutative torus can be identified with an element of the 
K-group $K_0$ of the noncommutative torus. We call the corresponding element $\mu(F)$. 
This can be seen as an element of the integral even cohomology $H^{even}(\T^d_\theta,\Z)$. 
On the classical torus $\mu(F)$ can be  identified with the Chern character. On the 
noncommutative torus this is no longer true. In fact, the Chern character can be expressed 
in terms of $\mu(F)$ as \cite{ell,rie}
\begin{equation}
\ch(F)=\e^{-\theta}\con\mu(F), 
\end{equation}
where $\con$ denotes contraction.
Note that $\theta$ should be considered a bivector, so  that it can be contracted with forms 
in a coordinate invariant way. This element is not integral any more. Note however that 
the top chern character $\int\! d^dx \ch(F)$ is still integral. This is because it can be 
related to an index, analogous to the classical torus \cite{con}. 

In two dimensions, the element $\mu(F)$ can be identified with 
\begin{equation}
\mu(F)=N+M dx^{12}. 
\end{equation}
Here we use a shorthand notation $dx^{12}=dx^1\wedge dx^2$. The Chern Character can 
then be written 
\begin{equation}
(N+M\theta) + M dx^{12}.
\end{equation}
We get immediately the normalization of the trace, 
$\Tr_\theta I=N+M\theta$. The magnetic flux, as it is the top Chern character 
in two dimensions, is still an integer. 

In four dimensions, the formulas get a bit more involved. The element $\mu(F)$ can be written 
\begin{equation}
\mu(F)=N+\frac{1}{2}M_{ij}dx^{ij}+k dx^{1234}.  
\end{equation}
The Chern character of this bundle takes the form 
\begin{equation}
\ch(F)=\Bigl(N+\frac{1}{2}\theta^{ij}M_{ij}+\frac{1}{2}\theta\wedge\theta k\Bigr) 
+ \frac{1}{2}(M_{ij}+\st \theta_{ij}k)dx^{ij}
+k dx^{1234}.
\end{equation}
Again, the top Chern character, which is now the instanton number $k$, is integral. 
From this we can read off the zero modes of the operators $I$, $F$ and $F\wedge F$. 
Note that for the gauge connection that we explicitly described earlier, 
the matrix $M_{ij}$ in the formulas above is equals the matrix of fluxes appearing 
in the field strength.

\subsection{Quantization of the Electric Flux}

We now study the periodicity of the gauge field. As the $\U(N)$ gauge field takes values in 
a compact space, there are certain periodicity conditions. Most important is the $\U(1)$ 
factor, which gives a periodicity $\CA_i\to \CA_i+2\pi$. This periodicity is generated 
by the electric field operator $E^i=\delta/\delta \CA_i$. On the commutative torus this 
shift is generated by a gauge transformation, with $\Omega=\exp 2\pi ix^i$. On the 
noncommutative torus this is not a local gauge transition function, as we saw above. The 
gauge transformation that is most close to this on the noncommutative torus is 
$\Omega=\exp(2\pi ix^i+\theta^{ij}\grad{j})$.\footnote{This corrects a similar expression 
for the gauge transformation in an earlier paper.} Note that this is a function of the 
$\tilde x^i$, if we combine the derivative in $\grad{}$ with the coordinates. Also note 
that this gauge transformation commutes with the $T_i$. We comment on the necessity for 
this below. We find 
\begin{equation}
\grad{j}\to\e^{\theta^{ik}\grad{k}}\grad{j}\e^{-\theta^{ik}\grad{k}}-2\pi i\delta_{ij}.
\end{equation}
Now the shift part is generated by the electric field zero mode $\int\!\Tr E^i$, while 
the covariant derivative generates a translation, which can be interpreted as the action of  
the total momentum operator, $\int\!\Tr_\theta P_i$. The gauge transformation on the gauge 
field above is therefore generated on the wave function by the operator 
\begin{equation}
\exp\Bigl(2\pi i\int\!\Tr_\theta\,(E^i+\theta^{ik}P_k)\Bigr).
\end{equation}
As it is a true gauge transformation ($\Omega$ is single valued on the noncommutative 
torus), this operation should act trivial on the wave function. Therefore, the 
quantization of the electric flux is modified on the noncommutative torus to 
\begin{equation}\label{quantEP}
\int\!\Tr_\theta E^i = n^i-\theta^{ij}m_j,\qquad \mbox{where}\qquad
\int\!\Tr_\theta P_i = m_i.
\end{equation}
Here both $n^i$ and $m_i$ are integers. 
Note that the total momentum is still quantized in the usual way, because they are related 
to the periodicity of the torus. 

The conclusions we derived above for the periodicity of the gauge field is different from 
the one calculated in \cite{bmz,brmo}. As remarked in the last paper, and as we shall see 
shortly in more detail, this has an important consequence for the BPS spectrum of gauge 
theories. It is therefore useful to comment in some more detail on what is going on. 
First of all, it is essential to identify the global gauge transformations correctly. 
Technically, they should be global sections of the principal bundle corresponding to the 
gauge bundle at hand. In the formulation that we have been using, this just means that they 
should commute with the translation operators $T_i$. We already argued that any gauge 
transformation (either global or local) is a function of the $\tilde x^i$. Therefore we 
find that \emph{global} gauge transformations are single-valued functions of the $Z_i$. 
For example the gauge transformation we have been using above satisfies this criterium, 
because both $U_i$ and $\grad{j}$ commute with the $T_i$. The global gauge transformations 
that were used in the two papers mentioned above, and which let them to conclude that the 
periodicity was different, where generated by the $Z_i$ themselves. As these are roughly 
proportional to $\exp 2\pi ix^i/(N+M\theta)$ it seems that the periodicity of the gauge field is 
proportional to $(N+M\theta)\inv$. Now how can we explain this 
discrepancy? The difference arises not because we use a different gauge transformation, 
but because the gauge-transformed gauge field should not directly be identified with the 
shifted gauge field. More careful inspection shows that it must be identified in general 
with a gauge field that is \emph{both shifted and translated}. The fact that also a translation 
should be considered can be seen if we write the full gauge connection $\grad[0]{i}+i\CA_i(Z)$. 
Acting with a global gauge transformation, for example $Z_i$, simply shifts the linear 
connection by $(N+M\theta)\inv$ (it could not do anything else). But this gauge 
transformation also effects the fluctuation part $\CA(Z)$. (This is in fact true also for 
the commutative torus). Now the commutation relations \eqref{commZ} show that $\log Z_i$ 
acts as a derivation on functions of $Z$. Therefore, the gauge transformation $Z_i$ acts as 
an exponentiated derivation, or a translation. To be precise, it can be shown that for the 
gauge bundles we have constructed above, the operator $Z_i$ acts on functions of $Z$ as 
\begin{equation}
Z_i f(Z) Z_i\inv = 
\e^{(\kappa+\lambda\theta)^{ij}\grad{j}} f(Z) \e^{-(\kappa+\lambda\theta)^{ij}\grad{j}}.
\end{equation}
If we now interpret the 
shift in the gauge field as  a result of the gauge transformation $Z_i$, including this 
translational contribution, we find that it is in fact generated by the operator 
\begin{equation}
\exp\Bigl(2\pi i\int\!\Tr_\theta\,\bigl(\lambda(E+\theta P) +\kappa P\bigr)\Bigr).
\end{equation}
Now because $\lambda$ and $\kappa$ are integral matrices by definition, this is in perfect 
agreement with the quantization for $E$ and $P$ that we derived above. Note that the 
translation over $\kappa$ is over an integral times the period of the torus. Therefore, it 
should automatically act trivially.

\subsection{Morita Equivalence}

We saw above that the modes $Z_i$ of the non-abelian gauge field generate an algebra 
\eqref{commZ} of the same type as the abelian modes $\widetilde U_i$. Therefore, we can 
identify the modes $Z_i$ with modes of an abelian gauge field. This identifies the 
non-abelian gauge field $\CA$ with an abelian gauge field $\widehat\CA$. The corresponding 
abelian gauge bundle lives on a different noncommutative torus. Comparing the commutation 
relations of the $Z_i$ with those of the $\widetilde U_i$, we see that this should be a 
torus with noncommutative parameter $\hat\theta$. So we find that the \emph{twisted non-abelian} 
gauge field $\CA$ on the non-commutative torus $\T^2_\theta$ can be identified with a 
\emph{trivial} gauge field $\widehat\CA$ on the dual torus $\widehat\T^2_{\hat\theta}$. 
The gauge group is the unbroken part of the gauge group $\U(N_0)$, where $N_0$ is defined in 
\ref{twist}. This relation is a special kind of \emph{Morita equivalence} \cite{con,sch}. 
Note that $\hat\theta$ is related by an integral 
fractional linear transformation to the old $\theta$. The matrix which 
induces this transformation is the $\SO(d,d;\Z)$ matrix 
\begin{equation}
S=\pmatrix{ \lambda&\kappa\cr -\mu&\nu }.
\end{equation}
This then should be the $\SO(d,d;\Z)$ duality transformation which maps 
the data $(N,M)$ to $(N_0,0)$, where $\U(N_0)$ is the unbroken part of 
the gauge group. Later in this section we shall see how $\SO(d,d;\Z)$ acts 
explicitly on the data of the gauge bundle when $d=4$.

The relation of the linear connection $\grad[0]{}$ to the corresponding 
abelian connection $\hatgrad[0]{i}=\widehat\partial_i$, which lives on a 
bundle on the dual torus $\widehat\T^d_{\hat\theta}$, can easily be found from the 
transformation of the coordinates, which gives $\widehat\partial=(\beta^t)\inv\partial$. 
The result is 
\begin{equation}
\hatgrad[0]{i}=(\nu-\mu\theta)_{ij}\grad[0]{j}+2\pi i(\mu-\nu\alpha)_{ij}x^j.
\end{equation}
Note that the scaling factor in front of the covariant derivative is gauge invariant, 
that is independent of the precise choice for $\alpha$. 

The transformation of the total connection should be given by essentially the same formula. 
Therefore the gauge field $\CA$ transforms with the matrix $\nu-\mu\theta$. 
The transformation of the field strength is now easily derived 
\begin{equation}\label{trafoF}
\widehat F= (\nu-\mu\theta)F(\nu-\mu\theta)^t + \mu(\nu-\mu\theta)^tI.
\end{equation}
Note that this relation is completely gauge invariant. The scaling of the field 
strength and the gauge field can be undone by a linear coordinate transformation. 
Also, the shift guarantees that the flux on the dual torus becomes zero, making 
the dual bundle trivial. 

From \eqref{trafoF} we can read off that under a general $\SO(d,d;\Z)$ duality 
transformation 
\begin{equation}
\pmatrix{A&B\cr C&D}\in\SO(d,d;\Z),
\end{equation}
(the fluctuation part of) the gauge field and the field 
strength transform as 
\begin{equation}\label{trafoAF}
\CA\to (C\theta+D)\CA,\qquad
F\to (C\theta+D)F(C\theta+D)^t+C(C\theta+D)^tI. 
\end{equation}
The transformation of the electric field $E^i=\delta\CL/\delta\CA_i$ can most easily 
be found from the requirement that $\Tr_\theta E^i\dot\CA_i$ should be invariant, and 
the fact that the trace transforms according to  
\begin{equation}\label{trafotr}
\Tr_\theta \to \sqrt{\det(C\theta+D)}\inv\Tr_\theta.
\end{equation}
Therefore the electric field has to transform as 
\begin{equation}\label{trafoE}
E\to \sqrt{\det(C\theta+D)}\,[(C\theta+D)^t]\inv E.
\end{equation}
Note that all these transformations are correct in any dimension.

The transformation rules for the field strength and electric fields can now be used 
to find the transformations of the integral fluxes. we saw before, that in the calculation 
of the fluxes, the electric flux mixes with the momentum. Indeed, from the transformations 
of the electric and magnetic fields, we derive the following transformation rules  
\begin{equation}
E+\theta P\to \sqrt{\det(C\theta+D)}\Bigl(A(E+\theta P) + BP\Bigr),\qquad
P\to \sqrt{\det(C\theta+D)}\Bigl(C(E+\theta P) + DP\Bigr).
\end{equation}
These rules imply that the electric and momentum fluxes $n^i$ and $m_i$ combine into 
a vector of the Morita group $\SO(d,d)$. Also the rank, magnetic flux and 
fluxes of higher exterior powers of the field strength get mixed as we saw before. The 
transformation of these fluxes is much more involved, and we will not perform this here. 
It turns out that they transform in a spinor of $\SO(d,d)$. As a hint towards this, 
note they transform as the even forms, which can be identified with a bispinor, of the 
$\SO(d)$ rotation group. Furthermore, the spinor of $\SO(d,d)$ can be identified with 
the bispinor of the $\SO(d)$ subgroup. 

A nice property of the above transformation rules of the fields, is that they are 
independent of the Lagrangian that is used, as long as it is invariant under 
Morita equivalence. So, as we shall discuss later, this is correct for the Born-Infeld 
theory on the noncommutative torus. Also, because of the invariance of the action 
shown in \cite{sch}, it is also correct for Yang-Mills theory. At least, 
if we assume the correct transformation rules for the parameters of the theory.

\subsection{Morita Equivalence in Four Dimensions}

In four dimensions, the Morita equivalences form an $\SO(4,4;\Z)$ group. We saw above 
that the noncommutative parameter $\theta$ transforms with fractional linear 
transformations under this group. As noted above, the fluxes of powers of the field 
strength transform in the spinor $\mathbf{8_s}$ of $\SO(4,4;\Z)$. These are the rank 
$N$ the magnetic fluxes  $M_{ij}$ and the instanton number $k$. It is useful to 
represent them into a matrix, which transforms in the $\mathbf{8_v}\otimes\mathbf{8_c}$ 
of $\SO(4,4;\Z)$.\footnote{Note that this product contains a $\mathbf{8_s}$ factor.} In 
\ref{so44}, we give an explicitly representation of the spinors, which we can use to 
identify this matrix. The electric flux and the total momentum transforms, as noted above, 
in the  vector $\mathbf{8_v}$ of the duality group. We find for this matrix and vector 
\begin{equation}\label{fluxes}
\CN=\pmatrix{N\one&\st M\cr M&-k\one},\qquad 
\CM=\pmatrix{n\cr m}.
\end{equation}
The explicit transformation rules are given by 
\begin{equation}
\CN\to \pm S\CN\widetilde S\inv,\qquad 
\CM\to S\CM
\end{equation}
where $S$ and $\widetilde S$ are the $\SO(4,4;\Z)$ transformations acting on the 
$\mathbf{8_v}$ and $\mathbf{8_c}$ respectively. The extra plus/minus sign is added 
to reflect the fact that the quantities are spinors. The sign should be chosen such 
that the rank $N$ is always positive. There is an invariant 
\begin{equation}\label{invar}
Nk-\frac{1}{2}M\wedge M.
\end{equation}
The invariance follows easily, because it can be related ot the determinant of $\CN$. 

The group $\SO(4,4;\Z)$ is generated by three types of elements. We shortly list 
these generators below, and discuss their action on the fluxes.

\begin{description}

\item[Mapping class group] These are just the basis change 
transformations of the four-torus. They are embedded in $\SO(4,4;\Z)$ as
\begin{equation}
S_A= \widetilde S_A= \pmatrix{(A^t)\inv&0\cr 0&A},
\end{equation}
where $A\in\SL(4,\Z)$ is the element of the mapping class group. 
It is easily seen that they leave $N$ and $k$ invariant, and transform $M_{ij}$ 
in the proper way as an antisymmetric matrix. Also the parameter $\theta$ 
transforms appropriately as an antisymmetric matrix.

\item[Integral shift transformations] They are generated by matrices 
of the form 
\begin{equation}\label{thshift}
S_\Theta = \pmatrix{ \one&0\cr \Theta&\one }, \qquad
\widetilde S_\Theta = \pmatrix{ \one&-\st \Theta\cr 0&\one },
\end{equation}
where $\Theta$ may be any antisymmetric matrix with integral entries. 
They shift the magnetic fluxes $M$ by an amount $N\Theta$. The transformation 
of the instanton number $k$ is precisely such that the combination \eqref{invar} 
is invariant. The parameter $\theta$ of the noncommutative torus transforms 
according to 
\begin{equation}
\theta\inv\to\theta\inv+\Theta.
\end{equation}

\item[Nahm transformations] These are generated by two-dimensional versions 
of the Nahm transformation. For example, the Nahm transformation in the 
12-direction is generated by the matrices 
\begin{equation}
S_{{12}} = 
  \pmatrix{ 0&0&\one&0\cr 0&\one&0&0\cr \one&0&0&0\cr 0&0&0&\one  },\qquad
\widetilde S_{{12}} = 
  \pmatrix{ \eps&0&0&0\cr 0&0&0&\eps\cr 0&0&\eps&0\cr 0&\eps&0&0  }, 
\end{equation}
where $\eps=i\sigma^2$ and $\one$ is the 2-by-2 unit matrix. 
The fluxes transform according to
\begin{equation}
N\to M_{12},\qquad k\to -M_{34},\qquad M_{12}\to -N,\qquad M_{34}\to k,\qquad 
M_{1i}\leftrightarrow \pm M_{2i}, 
\end{equation}
where $i=3,4$ in the last item. Note that it correctly interchanges the rank and 
the 12-component of the magnetic flux. The usual Nahm transformation, involving 
all four directions, is generated by the product 
$S_{1234}=-\widetilde S_{1234}=S_{12}S_{34}$. It interchanges the rank $N$ with 
the instanton number $k$, and replaces the magnetic flux $M$ by the negative of 
its Hodge-dual, $-\st M$. 

\end{description}

In the correspondence to string theory, Morita equivalence should be related to 
T-duality. The Nahm transformations are then related to the pure T-dualities. 
The careful reader may have noticed that there is no element corresponding to 
T-duality in a single direction. The reason for this, is that it takes us from 
the Type \II{A} strings to the Type \II{B} strings. Therefore, it is not an 
equivalence of the corresponding gauge theory, as this is only related to the 
Type \II{A} description. In fact, such a transformation is an element of 
$\O(4,4;\Z)$ and not of $\SO(4,4;\Z)$. Such a transformation interchanges  
magnetic fluxes with electric fluxes and momenta. This is the reason why they 
can not be seen so easily in this formulation.

\section{Born-Infeld on the Noncommutative Torus}
\label{ncbi}

In this section, we discuss Born-Infeld theory on the noncommutative four-torus 
as an application of the Morita equivalence. Our main interest will therefore be 
the construction of a manifestly invariant Hamiltonian for the Born-Infeld theory 
on the noncommutative four-torus.

\subsection{Lagrangian and Hamiltonian on the Classical Torus}

The correct effective description of D-branes, at least for vanishing Kalb-Ramond 
field,  has been known for quite some time \cite{lei}, and is described by a 
Born-Infeld action. This action for the abelian gauge theory is given by
\begin{equation}\label{actionbi}
S_{BI} = \frac{1}{\ell_s}\int\! d^4xdt\, 
            \frac{-1}{\lambda_s}\sqrt{\det\bigl(-G-\CF\bigr)}
            + \frac{1}{2}C_1\wedge\CF\wedge\CF + C_3\wedge\CF,
\end{equation}
where $\CF=F-B$.
The coupling constant in front is the ten-dimensional string coupling. Note that in 
Type \II{A} theory there are a one-form and a three-form RR field.

We can write the Hamiltonian density in a form which at first sight looks more 
invariant, by combining the various operators into a matrix and a vector 
\begin{equation}\label{EB}
\CB=\pmatrix{ I& \st F^{ij}\cr F_{ij}& -\frac{1}{2}F\wedge F },\qquad
\CE=\pmatrix{ E^i\cr P_i }.
\end{equation}
These are such that their zero-modes are precisely the matrix and vector of fluxes 
in \eqref{fluxes}. The momentum density is given by $P_i=F_{ij}E^j$. Using these 
combinations, we can express the Hamiltonian density into a form which looks already 
quite invariant 
\begin{equation}\label{hamil}
\CH = \frac{1}{\ell_{pl}}\sqrt{
   \frac{1}{4\lambda_6}\tr\Bigl(\bbG\CB\widetilde\bbG\CB^t\Bigr)
   + \lambda_6(\CE+\CB\bbC)^t\bbG(\CE+\CB\bbC) }.
\end{equation}
Here, the moduli are encoded into the following matrices 
\begin{equation}
\bbG=\pmatrix{ G-BG\inv B& BG\inv\cr -G\inv B& G\inv },\qquad
\bbC=\pmatrix{ \st C_3\cr -C_1-\st B\,\st C_3 },
\end{equation}
and $\widetilde\bbG$ is similar to $\bbG$, with $G$ replaced by $\widetilde G\inv=\sqrt{\det G}G\inv$, and 
$B$ by $\st B$. In order to have that at least the zero-mode contribution is invariant 
under $\SO(4,4;\Z)$, they should transform according to 
\begin{equation}
\bbG\to (S^t)\inv\bbG S\inv,\qquad 
\widetilde\bbG\to \widetilde S\widetilde\bbG\widetilde S^t,\qquad 
\bbC\to \widetilde S\bbC.
\end{equation}
These can be seen to give exactly the T-duality transformations for the moduli. 
The coupling is the effective coupling in the six-dimensional transverse world, given by 
$\lambda_6=\lambda_s/(\det G)^{1/4}$.

\subsection{BPS-Spectrum}

The BPS mass levels were 
calculated in \cite{hvz}, and also turn out to be given by a completely U-duality 
invariant expression. The result of that paper can be written in our notation in 
the form
\begin{eqnarray}\label{BPSmass}
\ell_{pl}^2M_{BPS}^2&=&\frac{1}{4\lambda_6}\tr\Bigl(\bbG\CN\widetilde\bbG\CN^t\Bigr)
   +\lambda_6\CM^t\bbG\CM\\ 
   &&+\sqrt{\CM^tJ\CN\widetilde\bbG\CN^tJ\CM+\frac{\lambda_6^2}{4}(\CM^tJ\CM)^2
   +\frac{1}{\lambda_6^2}\Bigl(\frac{1}{8}\tr(\CN^tJ\CN J)\Bigr)^2}.
   \nonumber
\end{eqnarray}
For notational simplicity we took $\bbC=0$. The manifest invarance under the 
$\SO(4,4;\Z)$ T-duality group is obvious from the transformation of tghe various 
components. 

Note that this result is 
certainly correct for $N=1$ and $b=0$, as we know that then the Born-Infeld theory 
is correct. For different values of $N$ and $b$, this expression is the unique 
expression for the BPS-mass which is invariant under the full expected U-duality
group $\E_5(\Z)=\SO(5,5;\Z)$. Therefore, whatever theory should describe the 
D0-branes, we know that it must always give rise to these BPS-masses. For a 
review on U-duality in the context of D-branes and M-theory, see \cite{obpi}. 

Also note that the contribution from the electric field is different from \cite{brmo}. 
The difference is a consequence of the correct periodicity of the gauge field $\CA$, 
and therefore the quantization of the electric field; as has been argued in the 
last section.

\subsection{Hamiltonian on the Noncommutative Torus}

On the noncommutative torus, the operators $\CB$ and $\CE$ can not be directly 
identified with the local fields as in \eqref{EB}. The reason for this is that with 
this identification, they have different zero-modes. We should therefore replace them 
with different ones, with the same integral zero-modes. First let us denote the 
right-hand-sides of \eqref{EB} on the noncommutative torus by $\CB_\theta$ and 
$\CE_\theta$. Denoting the zero modes of $\CB_\theta$ and $\CE_\theta$ by 
$\CN_\theta$ and $\CM_\theta$ respectively, we find that we can write 
\begin{equation}
\int\!\Tr_\theta\CB_\theta \equiv \CN_\theta = S_\theta^t\CN(\widetilde S_\theta^t)\inv,\qquad 
\int\!\Tr_\theta\CE_\theta \equiv \CM_\theta = S_\theta^t\CM. 
\end{equation}
where $\Tr_\theta$ denotes the trace for the bundle on the noncommutative torus, 
and $S_\theta$ and $\widetilde S_\theta$ are theta-shift matrices as in \eqref{thshift}. 
The same relation should then apply for the local operators. It was argued in \cite{codo} 
and other papers, that the noncommutative parameter $\theta$ should be identified in string 
theory compactifications with the T-dual $B$-field. After making this identification, the 
Hamiltonian can be written 
\begin{equation}\label{hamilnc}
\CH = \frac{1}{\ell_{pl}}\Tr_\theta\sqrt{
   \frac{1}{4\lambda_6}\tr\Bigl(\CG\CB_\theta\widetilde\CG\CB_\theta^t\Bigr)
   + \lambda_6(\CE_\theta+\CB_\theta\CC)^t\CG(\CE_\theta+\CB_\theta\CC) },
\end{equation}
where now the moduli are given by 
\begin{equation}
\CG=\pmatrix{ g\inv& 0\cr0& g },\qquad
\widetilde\CG=\pmatrix{ \tilde g& 0\cr0& \tilde g\inv },\qquad
\CC=\pmatrix{\st c_3\cr -c_1}.
\end{equation}
Here $g$ and $c$ are the metric and RR-fields in the T-dual picture. Note that the 
T-dual $B$-field only appears as the noncommutative parameter. Note that the appearance 
of the metric in the Hamiltonian is such that the metric in the gauge theory is 
identified with $g\inv$, which is the inverse of the metric the $N$ D0-branes see. 
This is exactly the metric appropriately for the dual momentum torus. 

When taking the trace in the above formula for the Hamiltonian density, one has to 
be careful about the ordering of the operators. It was argued that for the 
non-abelian generalization of the Born-Infeld, as it follows from string theory, 
one should take the symmetric trace \cite{tsey}. As we are ultimately trying to 
describe string theory, we shall also assume that the trace here is a symmetric trace. 
For the paper at hand, this will however not be very important.

\subsection{Manifest $\SO(4,4;\Z)$ Duality}

Now that we have constructed the theory on the noncommutative torus, we can start to 
analyze it. We already saw that we have the correct BPS mass levels, invariant even 
under the full U-duality group. 

As we already noted before, the local operators $\CE$ and $\CB$ should transform, 
modulo normalization, as a spinor and vector under this duality group, because the 
zero-modes of these operators do so. In order to find the correct transformation of 
the zero-modes, the local operators should transform according to the rule 
\begin{equation}\label{trafoEB}
\CB^\prime = d S\CB\widetilde S\inv, \qquad
\CE^\prime = d S\CE
\end{equation}
where $d=\Tr_\theta I/\Tr_{\theta^\prime}I$ takes care of the difference in normalization. 
In noncommutative geometry, it has the interpretation of the scaling of an abstract 
dimension of the fiber of the bundle over the noncommutative torus (which is not 
necessarily integral on the noncommutative torus). To appreciate this, note that for 
the classical torus it equals the scaling of the rank $N$. 

Let us see what this means for the fields on the noncommutative torus. Before we 
noted that problems arose because the unit operator transforms into a combination of 
the unit operator and the field strength. So let us try to find a transformation such 
that the transformation of the unit operator does not get additional contributions. It 
turns out that, with the transformation rule \eqref{trafoEB} and the relation of 
$\CE$ and $\CB$ to the fields on the noncommutative torus, the noncommutative 
parameter $\theta$ must transform with fractional linear transformations 
\begin{equation}\label{trafoth}
\theta^\prime = (A\theta+B)(C\theta+D)\inv.
\end{equation}
Using this, we can calculate the scaling of the trace from the formula for the trace of 
the unit
\begin{equation}
d = \frac{\Tr_\theta I}{\Tr_{\theta^\prime}I} = \sqrt{\det(C\theta+D)}. 
\end{equation}
Using the various relations, we find that in the end the objects $\CE_\theta$ and 
$\CB_\theta$ transform with the 
matrix 
\begin{equation}
S_{\theta^\prime}^tS(S_\theta^t)\inv 
= \pmatrix{\bigl((C\theta+D)^t\bigr)\inv&0\cr C&C\theta+D},\quad
\widetilde S_{\theta^\prime}^t\widetilde S(\widetilde S_\theta^t)\inv 
= \pmatrix{(\widetilde A\,\st\theta-\widetilde B)\inv&0\cr \widetilde A^t&\widetilde A^t\,\st\theta-\widetilde B^t}.
\end{equation}
Now we can extract the transformation of the field strength $F$ and the 
electric field $E$. They are exactly the transformation that is expected from 
Morita equivalence for bundles on the noncommutative torus. 
Furthermore, it turns out that in this limit the unit operator $I$ remains unchanged. 
This is a very important consistency check, because the unit operator has no 
fluctuations, while all the other local operators indeed have fluctuation contributions. 
Hence we find that the $\SO(4,4;\Z)$ Morita equivalence is a manifest symmetry of the 
Born-Infeld theory.

\section{Discussion and Conclusion}
\label{concl}

In this paper we gave some explicit constructions of gauge bundles on noncommutative tori. 
We saw that Morita equivalence is a very natural identification of sections of 
the adjoint bundles. We were able to construct bundles with magnetic fluxes. But 
for the bundles we could construct the higher topological characters, such as the 
instanton number, coming were completely from the abelian part. Fully non-abelian 
instantons on the torus, even the classical torus, are very hard to construct. But 
as we have seen, the construction of fluxes on the noncommutative tori is basically 
the same as for the classical tori. So this problem is not essentially due to the 
noncommutativeness of the geometry, but simply of our lack of understanding of 
these true non-abelian instantons. Related to this is the extra constraint on the 
magnetic fluxes, as discussed in \ref{twist}. This is due to the fact that the 
contribution from the non-abelian part and the abelian part of the gauge field to 
the instanton number should always add up to an integer, although they are not 
separately restricted to integer values. However, when the non-abelian 
contribution vanishes this gives -- at least in four dimensions -- exactly the 
restriction on the magnetic fluxes discussed in the appendix \cite{geba}. It would 
inded be very nice to give a construction also for these non-abelian instantons. 
But as we have  seen, the cases we could describe already give enough information on 
bundles transform under Morita equivalence. 

In the last section we saw that it is possible to write down a Born-Infeld 
action on the noncommutative four-torus which reproduces exactly the correct 
BPS-spectrum expected for the D4-brane. A subtlety arises in the identification 
of the noncommutative parameter $\theta$ with the dual $B$-field $b$, as they 
transform differently under the $\SO(d,d;\Z)$ T-duality transformations. Only in 
the limit of small volume $g\to0$ do their transformations match, and can we 
consistently identify these parameters. In \cite{hv} we argued that this is 
precisely the limit where we expect the Yang-Mills description to be correct. 
Note that although the volume itself should be zero, the scaling of the metric 
and the shape are still important in proving duality invariance. The identification 
of T-duality transformations with Morita equivalence between the bundles makes it 
possible to even leave the BPS-limit.In fact, noncommutative geometry gives a 
Born-Infeld theory which is fully invariant under these $\SO(d,d;\Z)$ transformation, 
precisely because it can be described by Morita equivalences. This is a definite 
improvement over the description using classical geometry, where this is not the 
case. Up till now nobody has been able to capture also the rest of the U-duality 
group $\E_{d+1}(\Z)$ as a manifest duality. Our BPS-spectrum up till $d=4$ is 
however still invariant under this group. IT would be interesting to see if somehow 
also these other dualities could be captured in terms of a noncommutative 
description of some kind.

\ack

It is a pleasure to thank G. Zwart, A. Criscuolo, P. van Baal and R. Dijkgraaf 
for interesting discussions and helpful comments. C.H. is  financially supported 
by the Stichting FOM. The research of E.V is partly supported by the Pionier 
Programme of the Netherlands Organisation for Scientific Research (NWO).

\appendix

\section{Twist-Eating Solutions}
\label{twist}

In the description of the quasi periodicity conditions for the fluxes we needed $\SU(N)$ 
matrices $V_i$ satisfying commutation relations 
\begin{equation}
V_iV_j = \e^{2\pi iM_{ij}/N}V_jV_i,\qquad i,j=1,\ldots,d,
\end{equation}
where $M_{ij}$ is an anti-symmetric integral matrix. We assume that $d=2g$. 
The solutions to these equations are 
called twist-eating solutions. There is a direct relation to these commutation 
relations and the Heisenberg algebra. This relation is the key to an explicit 
construction of these solutions, as described in \cite{geba}. We will only give 
the result of this construction here. 

First we write the fluxes in a canonical block-diagonal form
\begin{equation}\label{Mblock}
M = \pmatrix{M_1\eps\cr &M_2\eps\cr &&\ddots},
\end{equation}
using the $\SL(d,\Z)$ symmetry. So there are $g$ fluxes $M_i$. From now on we let $i,j,\ldots $ 
run from 1 to $g$. Introduce $n_i=\gcd(M_i,N)$ and $N_i=N/n_i$. It turns out \cite{geba} that twist eating 
solutions can only exists if $N_1\cdots N_g|N$. We write $N=N_1\cdots N_gN_0$. 
When this restriction is satisfied, an explicit construction of the twist-eating solution can 
be given by the tensor product of several $\SU(N_i)$ factors as
\begin{eqnarray}
V_{2i-1} &=& \one_{N_1}\otimes\cdots\otimes P_{N_i}\otimes\cdots\otimes\one_{N_g}\otimes\one_{N_0},\\
V_{2i} &=& \one_{N_1}\otimes\cdots\otimes (Q_{N_i})^{M_i/n_i}\otimes\cdots\otimes\one_{N_g}\otimes\one_{N_0},
\end{eqnarray}
where $P_N$ and $Q_N$ are $\SU(N)$ clock and shift matrices 
\begin{equation}\label{clockshift}
Q_N = \pmatrix{ ~1~\cr &\e^{\frac{2\pi i}{N}}\cr &&\e^{\frac{4\pi i}{N}}\cr &&&\ddots }, 
\qquad 
P_N = \pmatrix{ ~0~&~1~\cr &~0~&~1~\cr &&\ddots&\ddots\cr ~1~&&&~0~ }.
\end{equation}
The part of $\SU(N)$ commuting with all the matrices $V_i$ are obviously the 
$\SU(N_0)$ matrices of the form $\one_{N_1}\otimes\cdots\otimes\one_{N_g}\otimes V_0$. 
Hence the part of the gauge group $\U(N)$ left unbroken by this configuration of 
fluxes is $\U(N_0)$. (The $\U(1)$ part, as it is the center of $\U(N)$, also remains 
unbroken).

\section{$\SO(4,4)$ Clifford Algebra}
\label{so44}

The algebra of $\SO(4,4)$ has, just as for $\SO(8)$, three inequivalent 
eight-dimensional representations: a vector, a spinor and a conjugate spinor. 
In the following we will use 8-valued indices $I,J,\ldots$ for the vector, 
$A,B,\ldots$ for the spinor and $\dot A, \dot B,\ldots$ for the conjugate spinor. 

We shall construct Dirac matrices relative to the spinor representation of 
$\SO(4,4)$. Using triality the Dirac matrices relative to the other two 
representations can be found by permuting the indices. The index structure is 
$\Gamma^A_{I\dot B}$ and $\Gamma^A_{\dot AJ}$. 

To construct the Dirac matrices $\Gamma^A$, we start from Dirac matrices for 
$\SO(3,3)$, which will be embedded into the spinor representation. Hence 
for this embedding we have the decompositions 
$\mathbf{8_s}\to\mathbf{6}\oplus\mathbf{1}\oplus\mathbf{1}$ and 
$\mathbf{8_v},\mathbf{8_c}\to\mathbf{4_s}\oplus\mathbf{4_c}$. 
The $\SO(3,3)$ Dirac matrices will be denoted $\gamma^a$, where 
$a,b,\ldots=1,\ldots,6$ is the vector index for $\SO(3,3)$. 
They can be represented by the following 6 real matrices 
\begin{equation}
\gamma^a = \gamma^a_- = \eps\otimes\eta^a_-,\qquad
\gamma^{a+3} = \gamma^a_+ = \tau_1\otimes\eta^a_+\qquad 
a=1,2,3.
\end{equation}
Here the $\eta^a_\pm$ form a basis of (anti-) self-dual antisymmetric 
$4\times4$ matrices $*\eta^a_\pm=\pm\eta^a_\pm$. They can be represented by 
\begin{eqnarray}
\eta^a_+:&\qquad \epsilon\otimes\tau_1,\qquad 
\epsilon\otimes\tau_3,\qquad1\otimes\epsilon,\cr
\eta^a_-:&\qquad \tau_1\otimes\epsilon,\qquad 
\tau_3\otimes\epsilon,\qquad \epsilon\otimes1.
\end{eqnarray}
In all the formulas above we used the matrices 
\begin{equation}
\epsilon=\pmatrix{ 0&1\cr -1&0 },\qquad
\tau_1=\pmatrix{ 0&1\cr 1&0 },\qquad
\tau_3=\pmatrix{ 1&0\cr 0&-1 }.
\end{equation}
The chirality operator is 
$\gamma^7=\gamma^1\cdots\gamma^6=\tau_3\otimes\one$. 

The $\SO(4,4)$ Dirac matrices can now be represented by 
$\Gamma^A=\tau_1\otimes\gamma^A$ for $A=1,\ldots,7$, and 
$\Gamma^8=\eps\otimes\one$. Explicitly we have 
\begin{equation}
\Gamma^A=\pmatrix{ 0&\gamma^A_{I\dot B}\cr \gamma^A_{\dot AJ}&0 },
\qquad A=1,\ldots,7\quad\mbox{and}\qquad
\Gamma^8=\pmatrix{ 0&\delta_{I\dot B}\cr -\delta_{\dot AJ}&0 }.
\end{equation}
Let $\psi_A=(\psi_a,\psi_7,\psi_8)$ be a spinor in $\mathbf{8_s}$ (here $a$ runs from 1 to 6). 
We can use the explicit representation of the gamma-matrices to embed this spinor 
into $\mathbf{8_v}\otimes\mathbf{8_c}$, via
\begin{equation}
\Psi_{I\dot B}=\psi_A\gamma^A_{I\dot B}=
\pmatrix{ \psi_8+\psi_7&\psi_a\eta^a\cr \st (\psi_a\eta^a)&\psi_8-\psi_7 }
\end{equation} 

The generators of $\SO(4,4)$ on $\mathbf{8_v}\oplus\mathbf{8_c}$ are the 
matrices $\Gamma^{AB}$. They can be decomposed into diagonal blocks, acting 
on the two irreducible factors. We find 
\begin{equation}
\Gamma^{AB}=\one\otimes\gamma^{AB},\qquad
\Gamma^{8A}=\tau_3\otimes\gamma^A,\qquad 
A,B=1,\ldots,7. 
\end{equation}
From this we find the action on the vector representation through matrices  
\begin{equation}
\Gamma^{AB}_{IJ}=\gamma^{AB}_{IJ}.\qquad 
\Gamma^{8A}_{IJ}=\gamma^A_{IJ},\qquad 
A,B=1,\ldots,7, 
\end{equation}
and on the conjugate spinor by 
\begin{equation}
\Gamma^{AB}_{\dot A\dot B}=\gamma^{AB}_{\dot A\dot B}.\qquad 
\Gamma^{8A}_{\dot A\dot B}=-\gamma^A_{\dot A\dot B},\qquad 
A,B=1,\ldots,7.
\end{equation}

They all leave the quadratic form given by the matrix $J=\tau_1\otimes1\otimes1$ 
invariant. This means that $XJ+JX^t=0$ 
(both in the vector, spinor and conjugate spinor representation). 
This implies that the generators are of the form 
\begin{equation}
X = \pmatrix{ U+u\one&V\cr W&-U^t-u\one }, \qquad
\widetilde X = \pmatrix{ U-u\one&-\st W\cr -\st V&-U^t+u\one }.
\end{equation}
where $V$ and $W$ are anti-symmetric and $U$ is traceless. Here $X$ and $\widetilde X$ 
are the generators in the vector and the conjugate spinor representation respectively. 
The relatiion between the two can be found using the explicit relation beteween the 
two representations as described above.


\begin{thebibliography}{88}

\bibitem{codo} A. Connes, M.R. Douglas and A. Schwarz, 
\textit{Noncommutative Geometry and Matrix Theory: Compactification on Tori}, 
\texttt{hep-th/9711162}.

\bibitem{howu} P.-M. Ho, Y.-Y. Wu and Y.-S. Wu, 
\textit{Towards a Noncommutative Geometric Approach to Matrix Compactification},
\texttt{hep-th/9712201}.
\bibitem{howu2} P.-M. Ho and Y.-S. Wu, 
\textit{Noncommutative Gauge Theories in Matrix Theory},
\texttt{hep-th/9801147}.
\bibitem{dohu} M.R. Douglas, C. Hull, 
\textit{D-Branes and the Noncommutative Torus},
\texttt{hep-th/9711165}.
\bibitem{li} M. Li, 
\textit{Comments on Supersymmetric Yang-Mills Theory on a Noncommutative Torus},
\texttt{hep-th/9802052}.
\bibitem{cas} R. Casalbuoni,
\textit{Algebraic Treatment of Compactification on Noncommutative Tori},
\texttt{hep-th/9801170}
\bibitem{sch}A. Schwarz, 
\textit{Morita Equivalence and Duality}, 
\texttt{hep-th/9805034}. 

M. Rieffel and A. Schwarz, 
\textit{Morita Equivalence of Multidimensional Noncommutative Tori}, 
\texttt{q-alg/9803057}.
\bibitem{kaok} T. Kawano and K. Okuyama, 
\textit{Matrix Theory on Noncommutative Torus}, 
\texttt{hep-th/9803044}
\bibitem{chkr} Y.-K.E. Cheung and M. Krogh, 
\textit{Noncommutative Geometry from 0-Branes in a Background B-Field},
\texttt{hep-th/9803031}.
\bibitem{arar} F. Ardalan, H. Arfaei and M.M. Sheikh-Jabbari, 
\textit{Mixed Branes and M(atrix) Theory on Noncommutative Torus}, 
\texttt{hep-th/9803067}, and 
\textit{Noncommutative Geometry from Strings and Branes},
\texttt{hep-th/9810072}.
\bibitem{ho} P.-M. Ho, 
\textit{Twisted Bundle on Quantum Torus and BPS States in Matrix Theory}, 
\texttt{hep-th/9803166}.
\bibitem{mozu} B. Morariu and B. Zumino, 
\textit{Super Yang-Mills on the Noncommutative Torus}, 
\texttt{hep-th/9807198}.
\bibitem{bmz} D. Brace, B. Morariu and B. Zumino, 
\textit{Dualities of the Matrix Model from T-Duality of the Type II String}, 
\texttt{hep-th/9810099}
\bibitem{brmo} D. Brace and B. Morariu, 
\textit{A Note on the BPS Spectrum of the Matrix Model}, 
\texttt{hep-th/9810185}
\bibitem{hv} C. Hofman and E. Verlinde,
\textit{U-Duality of Born-Infeld on the Noncommutative Two-Torus}, 
\texttt{hep-th/9810116}.\bibitem{bfss} T. Banks, W. Fischler, S.H. Shenker and L. Susskind, 
\textit{M-Theory as a Matrix Model: A Conjecture}, 
Phys. Rev. \textbf{D55} (1997) 5112, \texttt{hep-th/9610043}.
\bibitem{tay} W. Taylor IV, \textit{D-Brane Field Theory on Compact Spaces}, 
Phys. Lett. \textbf{B394} (1997) 283, \texttt{hep-th/9611042}.

O.J. Ganor, S. Ramgoolam and W. Taylor IV, 
\textit{Branes, Fluxes and Duality in M(atrix) Theory}, 
Nucl. Phys. \textbf{B492} (1997) 191, \texttt{hep-th/9611202}.
\bibitem{lei}R. Leigh,
\textit{Dirac-Born-Infeld Action from Dirichlet Sigma Models}, 
Mod. Phys. Lett. \textbf{A4} (1989) 2767.
\bibitem{con} A. Connes,
\textit{Noncommutative Geometry}, 
Academic Press, 1994.
\bibitem{ell} Elliott, 
\textit{On the K-Theory of the $C^*$-Algebra Generated by a Projective 
Representation of a Torsion-Free Discrete Abelian Group}, 
Operator Algebras and Group Representations, Vol. 1, 
Pitman, London, 1984 159
\bibitem{rie} M. Rieffel, 
\textit{Projective Modules over Higher-Dimensional Noncommutative Tori}, 
Can. J. Math., Vol. XL, No. 2 (1988) 257.
\bibitem{geba} B. van Geemen and P. van Baal, 
\textit{The Group Theory of Twist Eating Solutions}, 
Proc. K. Ned. Akad. Wet., Ser. B, \textbf{89}, no. 1 (1986) 39.

B. van Geemen and P. van Baal, 
\textit{A Simple Construction of Twist Eating Solutions}, 
J. Math. Phys. \textbf{27}, no. 2 (1986) 455.  
\bibitem{hvz} C. Hofman, E. Verlinde and G. Zwart, 
\textit{U-Duality Invariance of the Four-Dimensional Born-Infeld Theory}, 
\texttt{hep-th/9808128}.
 \bibitem{obpi} N.A. Obers and B. Pioline, 
\textit{U-duality and M-Theory}, 
\texttt{hep-th/9809039}.
\bibitem{tsey}A.A. Tseytlin,
\textit{On Non-Abelian Generalization of Born-Infeld Action in String Theory},
Nucl.Phys. \textbf{B501} (1997) 41, \texttt{hep-th/9701125}.

\end{thebibliography}
\end{document}